# How to Document Computer Networks

**Sabah Al-Fedaghi and Bader Behbehani**

*Department of Computer Engineering, Kuwait University, Kuwait City, Kuwait*

**Abstract:** Documenting networks is an essential tool for troubleshooting network problems. The documentation details a network's structure and context, serves as a reference and makes network management more effective. Complex network diagrams are hard to document and maintain and are not guaranteed to reflect reality. They contain many superficial icons (e.g., wall, screen and tower). Defining a single coherent network architecture and topology, similar to engineering schematics, has received great interest. We propose a fundamental approach for methodically specifying a network architecture using a diagramming method to conceptualize the network's structure. The method is called a thinging (abstract) machine, through which the network world is viewed as a single unifying element called the thing/machine (thimac), providing the ontology for modeling the network. To test its viability, the thinging-machine-based methodology was applied to an existing computer network to produce a single integrated, diagrammatic representation that incorporates communication, software and hardware. The resultant description shows a viable, coherent depiction that can replace the current methods.

## Introduction

Computer networks are physical systems that form the backbone of modern technological societies. They are typically described using diagrams, which are a useful source of information when upgrading and modifying the architecture and presenting the networking environment to owners and decision-makers within an organization (Sopwith, 2019). Such diagrams have been used as tools to share a network's layout, "because the visual presentation makes it easier for users to understand how items are connected" (SmartDraw, 2019).

Documents are an important aspect of computer networks because of the strategic role of their information. According to Farmer *et al.* (2016), "The network documentation is a living representation of the design and implementation of the network equipment and its configuration". Documented network diagrams are "your best friends when you need to troubleshoot a problem" (Sherman, 2019). Troubleshooting involves a systematic approach that requires a thorough diagnosis of a network's characteristics, capabilities and history. In this context, documentation details a network's structure and context, serves as a reference and makes network management more productive (Portal Systems, 2019). Documentation provides details about the software, hardware, security and access, as well as invaluable functional information about the network. The key to finding an efficient solution for network problems is the ability to identify every element between the application's host server and the end user device that is accessing it (Sopwith, 2019). At a minimum, the network documentation should include the network device model, manufacturer, host name, serial number and location. We also need a drawing of the individual rack, a description of the equipment in the rack and its placement and even photographs of the network equipment (Sherman, 2019).

*Problem: Symptoms and Diagnostics*

Computer networks typically involve cross-organizational processes that "do not appear on the organizational map", which results in difficulties in creating process documentation (Ene and Persson, 2005). Organizations may decide to model these processes, but significant documentation problems arise because of their informal nature (Giagilis *et al.*, 1996). According to Sherman (2019), network administrators need to organize the network information and have the ability to search within it. We can use spreadsheets and shortcuts, such as a Wikis or even SharePoint (Sherman, 2019). Nevertheless, network diagrams are hard to maintain and keep current and they are not guaranteed to reflect reality (Kinghorn, 2018). According to Sopwith (2019), network diagrams are often created but rarely updated again until needed. Hence, a network summary that includes the following is necessary (Sopwith, 2019):

- The devices on the network (switches, firewalls, servers and important workstations)
- Network addresses (IP address ranges and subnet masks)
- Network services (DNS server addresses and ranges)
- Quality of service and high-level routing, where applicable
- Links to relevant procedures

Current depictions of network diagrams that have been developed over many years may include hundreds of different symbols, ranging from walls to computer screens to server racks to a cloud-based storage system, or may be based on an abstract graph theory representation that views a network as a set of nodes and edges. Scholars have been greatly interested in defining a single coherent representation of network architecture (Wolf *et al.*, 2012). This requires a rich understanding of relationships between elements of the network according to their applicability, environment and internal functionalities.

Our purpose with this paper is to create a diagrammatic language to specify the documentation of a computer network. The resultant description would provide a backbone for collecting all relevant network information. This language has been applied to the network of an actual organization (the second author's workplace) but is also generally suitable for other similar organizations.

*Proposed Approach*

We propose a more fundamental approach to specify network architecture methodically. We adopted a diagramming method from software engineering to help conceptualize the structure of network diagrams and the resultant schemata is used as a vehicle for documentation and communication among engineers, managers and decision makers. The method is called a Thinging (abstract) Machine (TM), through which the network world is viewed as a single unifying element that the thing/machine (thimac) uses to provide ontology for modeling network architecture. The viability of a TM-based methodology is demonstrated in a case study in which a single integrated, diagrammatic representation that incorporates communication, software and hardware is produced.

There are many types of computer networks (Yejian Technologies, 2020) and a network diagram may include hundreds of different symbols (SmartDraw, 2019) in different application fields of networking. UML diagrams are also used, even though no UML standard has specific diagrams to describe network architecture, but deployment diagrams could be used for this purpose with some extra networking stereotypes. UML's standard for a node or device is a three-dimensional view of a cube. In UML, the multiplicity of diagrams is a known problem (Dori, 2002), which contrasts with providing a single integrated, diagrammatic representation that incorporates function, structure and behavior as the TM does.

Olzak (2006) developed an architectural diagram to investigate logical data flow embedded in a network system. The method used to develop the network diagram apparently starts with a hardware diagram, followed by inspection of the diagram for possible logical connections between hardware components. Olzak (2006) also produced a functionality diagram that uses a data flow diagram approach to show the functional relationships between the various system components. Recently, Abdullah (2017) proposed representing computer networks using Petri nets as a tool that offers a diagrammatical modeling framework for the network design.

*Overview of the Paper*

Section 2 reviews the TM that was previously used in several published papers (Al-Fedaghi, 2019a-c; Al-Fedaghi and Makdessi, 2019; Al-Fedaghi and Alsumait, 2019; Al-Fedaghi and Al-Fadhli, 2019; Al-Fedaghi and BehBehani, 2018; Al-Fedaghi and Haidar, 2019; Al-Fedaghi and Al-Otaibi, 2019). The remaining sections describe applying TM to document an actual computer network in the case study.

## Thinging Machines

We adopt a conceptual process-based model centered on *things* and *machines* to model network systems. The philosophical foundation of this approach is based on Heidegger's notion of thinging (Heidegger, 1975). As described by Bryant (2012), "A tree is a thing through which flows of sunlight, water, carbon dioxide, minerals in the soil, etc., flow. Through a series of operations, the machine transforms those flows of matter, those other machines that pass through it, into various sorts of cells." According to Hassan *et al*. (2018), philosophy is not only intrinsically important, but it can also stand up in terms of some of the more established research metrics to other types of information systems research. Riemer *et al*. (2013) think that Heidegger's philosophy gives an alternative analysis of eliciting knowledge of routine activities, capturing knowledge from domain experts and representing organizational reality in authentic ways (Riemer *et al*., 2013).

Building on such an approach, things are unified with the concept of a process by being viewed as single ontological things/machines, or thimacs, which populate a world. A unit in such a universe has a dual purpose as a thing and as a machine.

A thing is a machine. Composite things are complex meshes of interwoven, interacting machines. The simplest type of thimac is called a TM, as shown in Fig. 1. The flow of things in a TM refers to the conceptual movement among five operations (stages). The stages of the TM can be described as follows.

**Arrive:** A thing flows to a new machine (e.g., packets arrive at a buffer in a router).

**Accept**: A thing enters a TM. For simplification, we assume that all arriving things are accepted; hence, we can combine arrive and accept as the **receiving** stage.

**Release**: A thing is marked as ready to be transferred outside the machine (e.g., in an airport, passengers wait to board after passport clearance).

**Process (change)**: A thing changes its form but not its "identity" (e.g., a node in the network machine processes a packet to decide where to forward it).

**Create**: A new thing is born in a machine (e.g., a logic deduction system deduces a conclusion).

**Transfer**: A thing is inputted into or outputted from a machine.

The TM includes one additional notation: Triggering (denoted by a dashed arrow in this paper's figures), which initiates a flow from one machine to another. Multiple machines can interact with each other through flows or triggering stages. Triggering is a transformation from one flow to another (e.g., a flow of electricity triggers a flow of air).

The thesis that *things are machines and machines are things* gives us a tool for handling things as processes. Thus, instead of the notions of class, attributes and methods in object-oriented methodology, TM has processes (machines) and sub-processes (submachines).

*Case Study of Documentation*

Our case study involves a single organization (the second author's workplace). The organization's business requirements are growing, causing an increase of service demands for Information Technology (IT) resources.

The IT department has acquired large number of servers and more storage capacity. Creating well-defined and understandable network documentation is a way to visualize an organization's network structure and is critical for improving the efficiency, effectiveness and timeliness of maintenance activities. This motivates the development of a TM-based model instead of using symbols such as walls, towers and human and computer icons, which do not produce systematic depictions that define coherent network architecture.

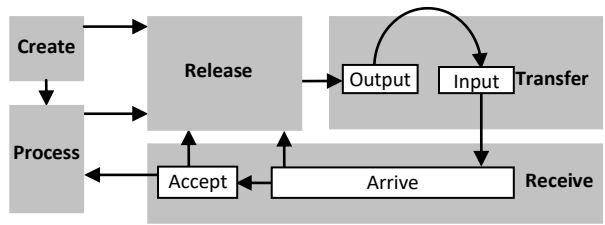

**Fig. 1:** A Thinging machine

Additionally, using abstract network architecture diagrams (graphs) of nodes and lines whose focus is solely to present the communication between the nodes is equally unsatisfactory because their extremely abstract content does not expose some of the nodes' internal functionalities. The individual features of the nodes and static and dynamic aspects of the graph are totally absent. In this case, the TM can form the foundation for a general description of the topological connectivity within the network that can be utilized to understand the network, the communication among different types of stakeholders, the maintenance process, monitoring and documentation.

In pursuing this goal, we will model packet flows through different sub-networks as the single phenomenon that ties different components of the network system together. TM modeling views the network as a thimac. To model the network as an existing system and as the result of inspecting the current network, we divide the TM thimac into several subthimacs (Fig. 2).

In the figure, Part A (Adaptive Security Appliance [ASA] and core switch – see descriptions later) describes the network portion of the communication flow from (i) the Internet and (ii) the WAN/LAN that connects the internal network. The detailed TM model of this part will be introduced in section 4.

Part B models the communication flow from internal sources (in the organization) that link to the common network. The detailed TM model of this part will be introduced in section 5.

All traffic coming or going from A and B are connected to Part C of the network that contains the application servers. The detailed TM model of this part will be introduced in section 6.

*Part A: Network Architecture of the Internet Communication*

The ASA (Fig. 2) is a basic component in many systems that provides users with highly secure access to data and network resources and delivers enterprise firewall capabilities. Figure 3, from Ryom (2019), shows a sample system that involves an ASA.

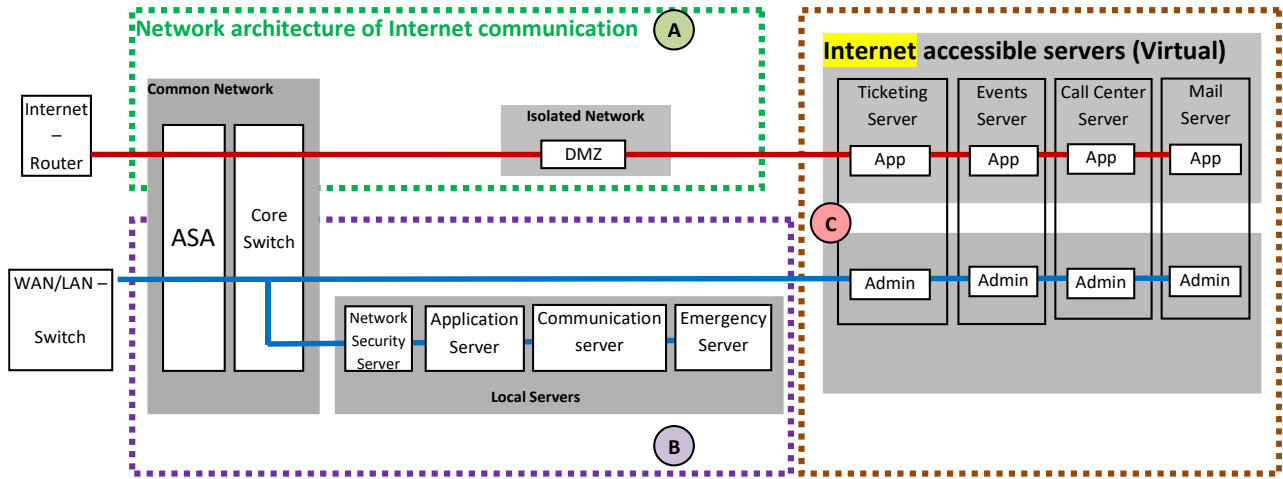

**Fig. 2:** A general description of the case study network

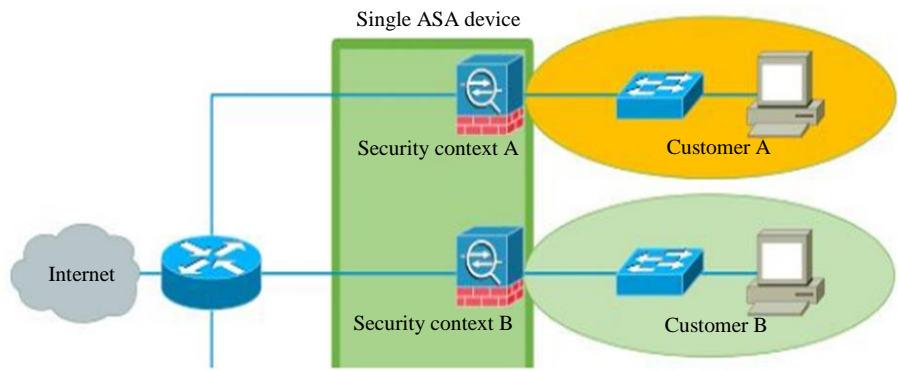

**Fig. 3:** A system that involves ASA (redrawn, partially from Ryom (2019))

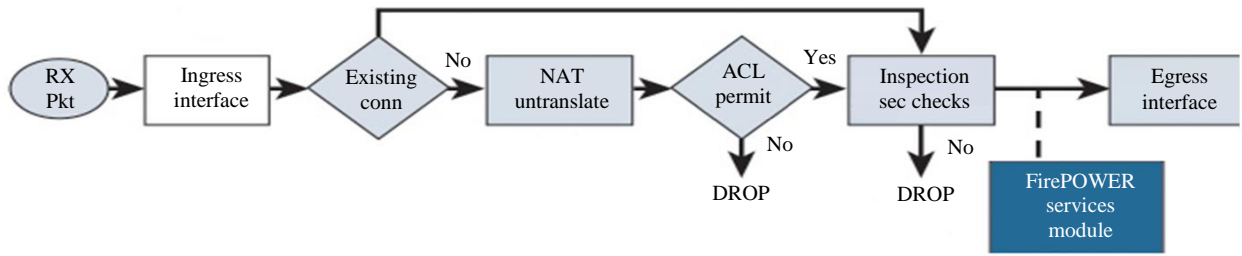

**Fig. 4:** ASA packet process algorithm (adopted from ASA 5500-X manual (2015))

**a.** The packet reaches the ingress interface.
**b.** Once the packet reaches the interface's internal buffer, the input counter of the interface is incremented by one.
Cisco ASA first looks at its internal connection table's details in order to verify if this is a current connection.
If the packet flow matches a current connection, then the access control list check is bypassed, and the packet is moved forward.
If the packet flow does not match a current connection, then the TCP state is verified.
If it is not a SYN packet, the packet is dropped, and the event is logged.
If it is a SYN packet or a User Datagram Protocol (UDP) packet, then the connection counter is incremented by one, and the packet is sent for an ACL check.
**c.** The packet is processed as per the interface ACLs.
If it matches any of the ACL entries, it moves forward. **…**

**Fig. 5:** Sample description of the ASA packet process algorithm (adopted from ASA 5500-X manual (2015))

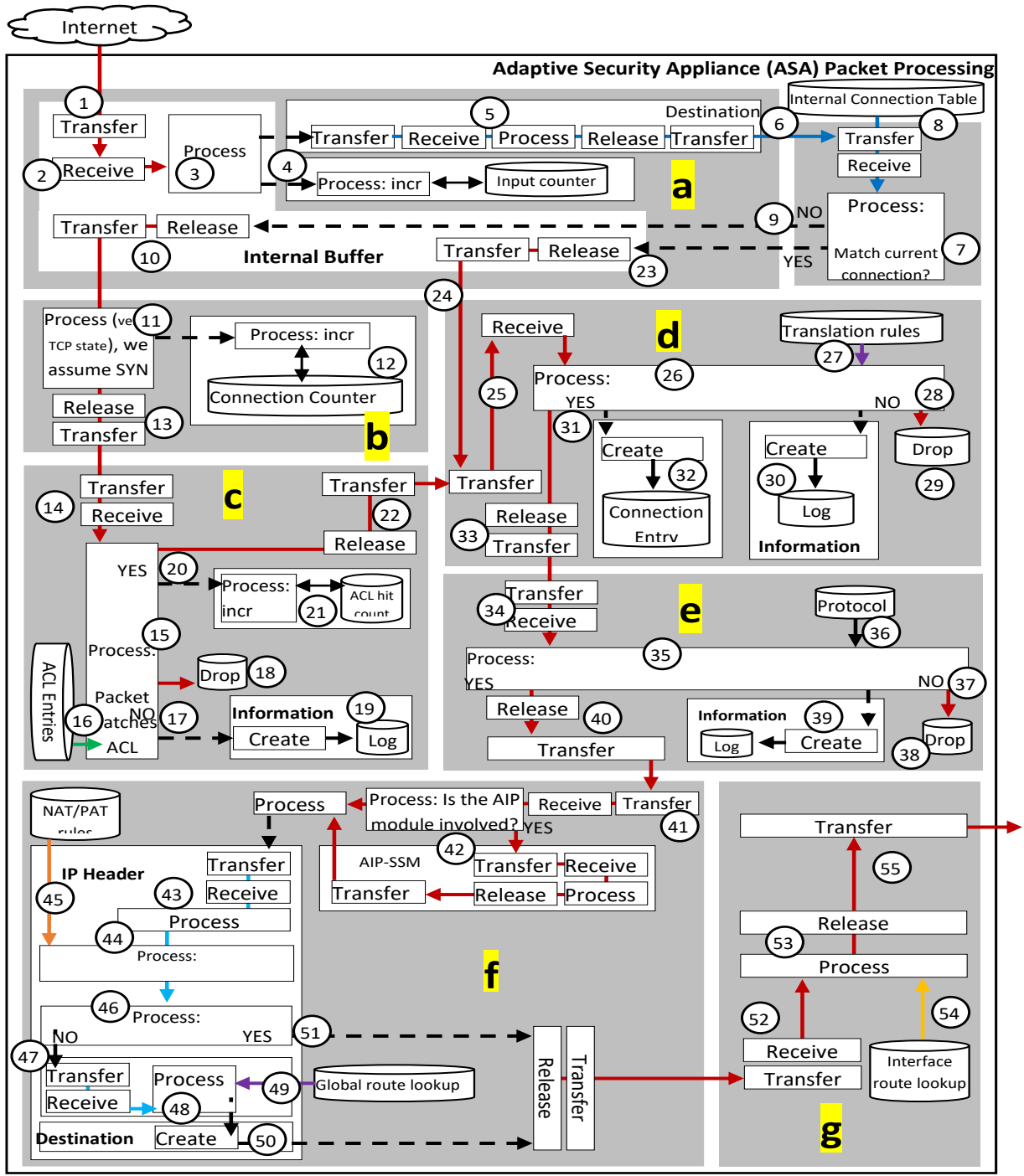

**Fig. 6:** The TM model of the ASA

Additionally, the ASA 5500-X manual (Cisco, 2015) includes a flowchart of the ASA packet process algorithm, as shown in Fig. 4 and described, partially, in Fig. 5.

*TM Model of the ASA*

Figure 6 shows the TM model of the ASA as described in the given sources. Due to space limitations, this TM model will not include certain parts, such as the firepower services module (Fig. 4).

Note that our TM description of the ASA is based on a user's (the second author's) understanding of the ASA's interior functionalities, which may not be complete or very accurate. However, the aim here is to demonstrate the type of logical specification using TM diagrammatic language, in contrast to superficial general diagrams such as that in Fig. 3.

If there is any inaccuracy, the TM diagram can easily be modified accordingly. Additionally, we can use the TM diagram to describe the architecture of the ASA to any desired level of detail, reaching the electronics level, depending on the purpose of the resultant TM specification (e.g., training or maintenance).

In Fig. 6, the packet flows from the Internet (through the router) as follows:

a.  The packet is first received in this part of the ASA.

    - The packet reaches the ingress interface (Circle 1), where it is received (2) and processed (3) as follows
    - The interface's input counter is incremented by one (4)
    - The internal connection is determined by extracting the destination from the packet (5) and flows (6), which are compared (7) with entries inside the internal connection Table (8) to verify if this is a current connection

b.  At this point of the ASA, the flow depends on the result of the process. If the result does not match the current connection (9), then the packet is released and transferred (10) to be processed with the TCP state verification (11) and the connection counter is incremented (12). We will assume that the packet is of type SYN to simplify (without loss of generality) the diagram's details. The packet is released and transferred (13) for an access control list (ACL) check

c.  At this point of the ASA, the packet is received (14) and compared (15) with ACL entries (16). If the packet does not match any ACL entries (17), then it is dropped (18) and the packet's information is logged (19). Otherwise (20), the ACL hit counter is incremented (21) and the packet is released and transferred (22)

d.  At this point of the ASA, the following occurs:

    - If the packet matches the current connection (23), then the packet is subjected to a verification process with translation rules (24) without passing through the TCP verification process (11) or an ACL check (14)
    - The packet is received (25) and verified (26) with translation rules (27). If the packet is not verified (28), then it is dropped (29) and the packet's information is logged (30). Otherwise (31), a new connection entry is created (32) and the packet is released and transferred (33)

e.  At this point of the ASA, the packet is received (34) and verified (35) if the packet flow complies with the protocol (36). If the packet does not comply with the protocol (37), then it is dropped (38) and the packet's information is logged (39). Otherwise, the packet is released and transferred (40)

f.  At this point of the ASA, the packet is received (41) and processed (42) with the Advanced Inspection and Prevention Security Services Module (AIP-SSM), if it exists. If it does not exist, the header is extracted (43) and translated (44) as per Network Address Translation/Port Address Translation (NAT/PAT) rules (45). The header is further processed (46) to check for an egress interface

    - If no egress is specified (47), then the destination is extracted (48) and updated according to global route lookup (49). Then, the packet is released and transferred (50) to the egress interface
    - If the egress is specified, then the packet is released and immediately transferred (51)

**g.** At this point of the ASA, the packet is received in the egress interface (52) and processed (53) with interface route lookup (54). Finally, the packet is released and transferred (55) out of the ASA

### Core Switch and Demilitarized Zone

Figure 7 shows the continuation of the packet flow through the core switch and the Demilitarized Zone (DMZ; Fig. 2. A general view of the flow destination is added to give a sense of where the packet flow is heading.

### Core Switch (Circles 41-55)

The core switch receives the packet (41) and proceeds as follows:

- The packet is processed for error detection (42)
- The packet is processed (43) by extracting its header, payload and trailer. The header fields consist of many pieces of information, such as the packet's total length, identification, fragment offset, time to live, protocol, header checksum, source address and destination address
- Each field in the header is processed (44) to produce (45) a new header for generating a new packet
- The payload or actual data (46) is processed to determine if the length is fixed (47). If so, then the data is padded (48) with random data to fill the empty gap. Otherwise (49), the packet is ready to be released and transferred (50)
- The trailer or footer (51) is processed (52) with an error-checking mechanism called the cyclic redundancy check (53). The newly created packet (54) is transferred (**55**) to DMZ

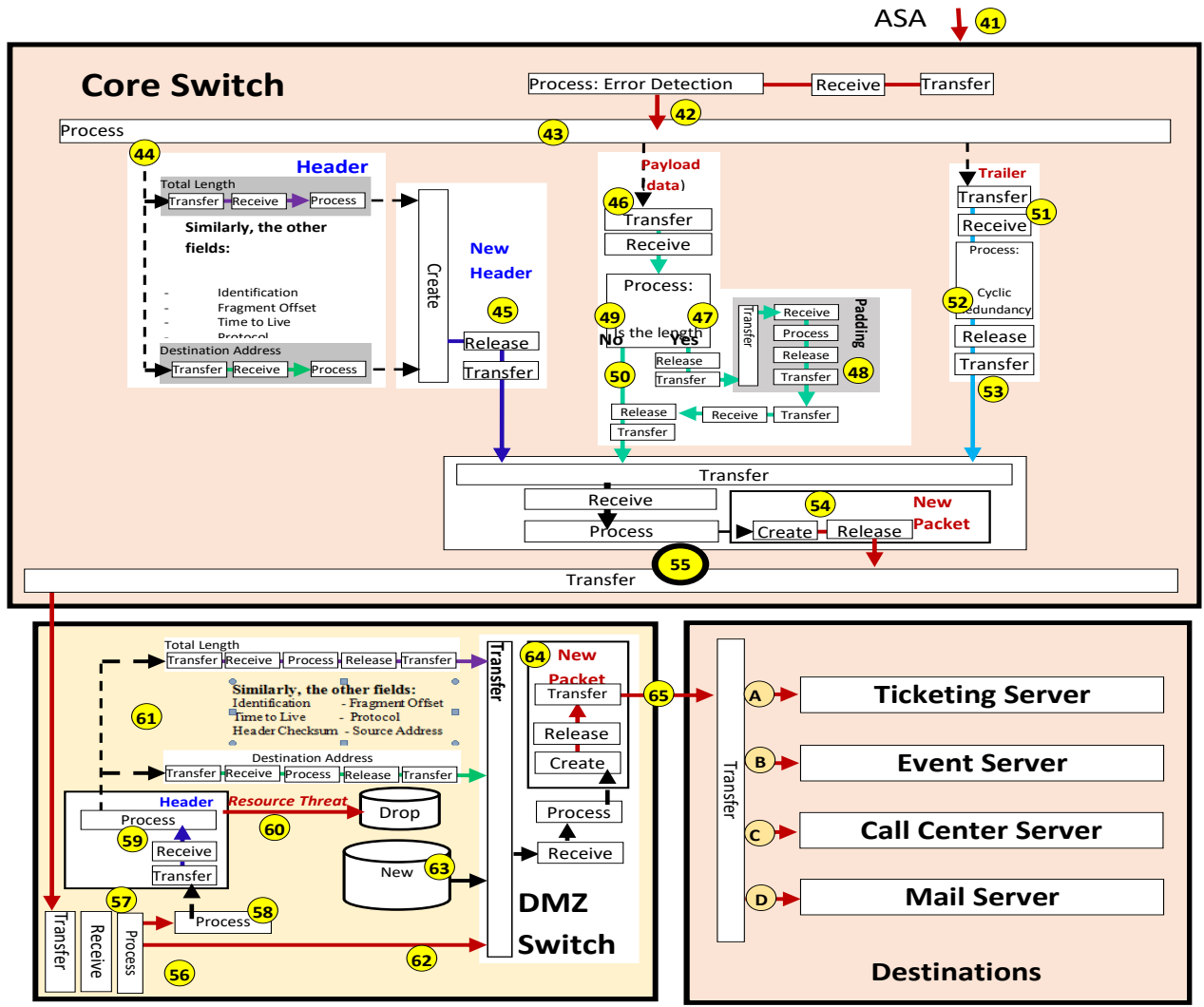

**Fig. 7:** Network architecture of Internet communication

*DMZ*

The packet flows to the DMZ (56), where it is processed (57) to produce the following:

(i). The packet is processed to extract the header, which, in turn, is processed to either (58)

- Drop the packet because of a threat (60) or
- Extract all header fields (61)

(ii). A new packet is constructed from the fields in the remaining parts of the old packet (62) and the new destination (63)

The new packet (64) flows to the destination server requested by the source (65). The packet might flow to one of the following servers:

(A) The ticketing server
(B) The event registration server
(C) The call center server
(D) The mail server

*Part B: Network Architecture of the Internal Communication*

Figure 8 shows the TM model of portion B of the network. Internal sources make requests to access one of the following types of local servers:

*Network Security Servers*

The packet flows to the network security server (66), where it is processed (67) to flow to one of the following (68):

i. The packet flows to the active directory server (69) to check (70) the requester's privileges and decide whether to drop the request (71) or to allow it (72) to proceed to the user station (the requester; 73)

The packet comes from the LAN for access permission (e.g., elevator access or office lock access). The server receives (74) the packet and decides (75) whether to reject the request (76) or allow it to proceed (77). The packet will be released and transferred to the requester (78) with an acknowledgment that access is granted or rejected

ii. The packet flows to the CCTV server for monitoring purposes. The CCTV server will receive (79) the packet and check its permissions (80) and upon a decision being made, the packet might be dropped (81) or allowed (82) to proceed. Then, it will be released and transferred to the requester (83) with an acknowledgment of the decision.

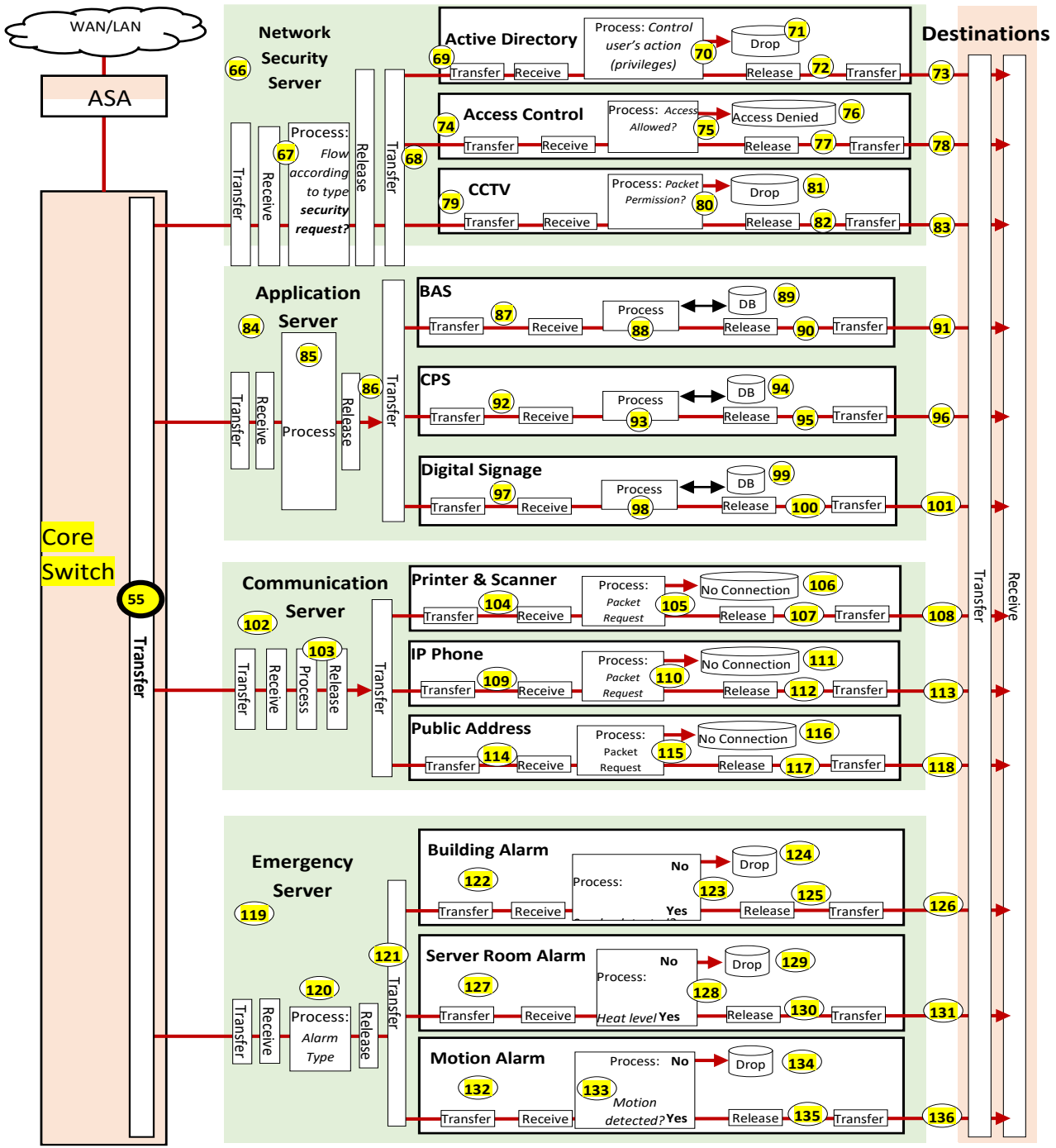

**Fig. 8:** Network architecture of internal communication

*Application Serve*

The core switch might forward the request to the application server if the request consists of requirements for accessing institutional systems such as the building automation system, car park system, or digital signage system. The application server will receive (84) the request and process (85) the requested system. Depending on the destination, the application server will transfer (86) this request to the specific system within the server:

- If the request reached the building automation system, then it will be received (87) and processed (88) with automation services (89), such as when a movement is detected. Then, the system will transfer (90) the result back to the requester (91)
- If the request reaches the car park system, then it will be received (92) and processed (93) to either update the data in the database or retrieve (94) data from it. Then, the system will transfer (95) the result back to the requester (96)
- If the request reaches the digital signage system, then it will be received (97) and processed (98) by either writing in the system or retrieving (99) from it. Then, the system will transfer (100) the acknowledgment back to the requester (101)

*Communication Server*

If the request is meant to reach a specific component in the organization (e.g., a printer, scanner, IP phone or public address system), then the communication server will receive (102) the request from the core switch. The server will start processing (103) the request by determining its destination:

- If the request is to achieve a printing or scanning process, then the packet will be transferred (104) to the printer or scanner, which will receive and process (105) the request. If the requester has no connection with the printer or scanner, then the request will be dropped (106). Otherwise, the request will be released and transferred (107) to the requester (108)
- If the packet requests access to a specific IP extension, then the packet will be transferred (109) to the corresponding location of the IP extension, which will receive and process (110) the request for acceptance. If the connection is lost, then the request will be dropped (111). Otherwise, the request will be released and transferred (112) to the respective destination with an acknowledgment (113)
- If the request is to reach the public address system, then the packet will be transferred (114) to the public address server, which will receive and process (115) the request for acceptance. If the connection with the public address system is lost, then the request will be dropped (116). Otherwise, the request will be released and transferred (117) to the requester (118)

*Emergency Server*

An emergency system can be a possible channel to receive (119) a request from the core switch. It will start processing (120) the packet's type to transfer (121) the packet to the proper channel:

- If the request is of the building alarm type, then the packet will be received (122) by the building alarm server, which will start the smoke-detection process (123). If no smoke is detected, then the packet will be dropped (124). Otherwise, the request will be released and transferred (125) to start the alarm (126)
- If the request is of the server-room alarm type, then the packet will be received (127) and processed (128) with heat-level detection. If no heat is detected, then the packet will be dropped (129). Otherwise, the request will be released and transferred (130) to start the alarm (131)
- If the request is of a motion alarm type, then the packet will be received (132) by the motion alarm switch. The motion alarm will start the motion-detection process (133). If no motion is detected, then the packet will be dropped (134). Otherwise, the request will be released and transferred (135) to start the alarm (136)

*Part C: Network Architecture of the Internet/Internal Communication*

Figure 9 shows the TM model of portion C of the network.

*Ticketing Server*

- For an Internet source (137), the packet flows from the DMZ to the ticketing server. Because the packet is transferred from the DMZ, the packet arrives in the ticketing application (139), where it is processed (140) to check the availability of the tickets. If the tickets are sold out, the packet will be dropped (141). Otherwise, the packet will be processed (142) for booking. The packet holding the booking information is stored (143) in the database. Then, the packet is released and transferred (144) back to the requester (145)
- For an internal source (138), the packet flows from the core switch to the ticketing server. Because the packet is transferred from the core switch, the packet arrives in the ticketing admin (146), where it is processed (147) to manage tickets, such as for changing details, deleting bookings and adding new prices. The new information is stored (148) in the database. Then, the packet is released and transferred (149) back to the requester (150)

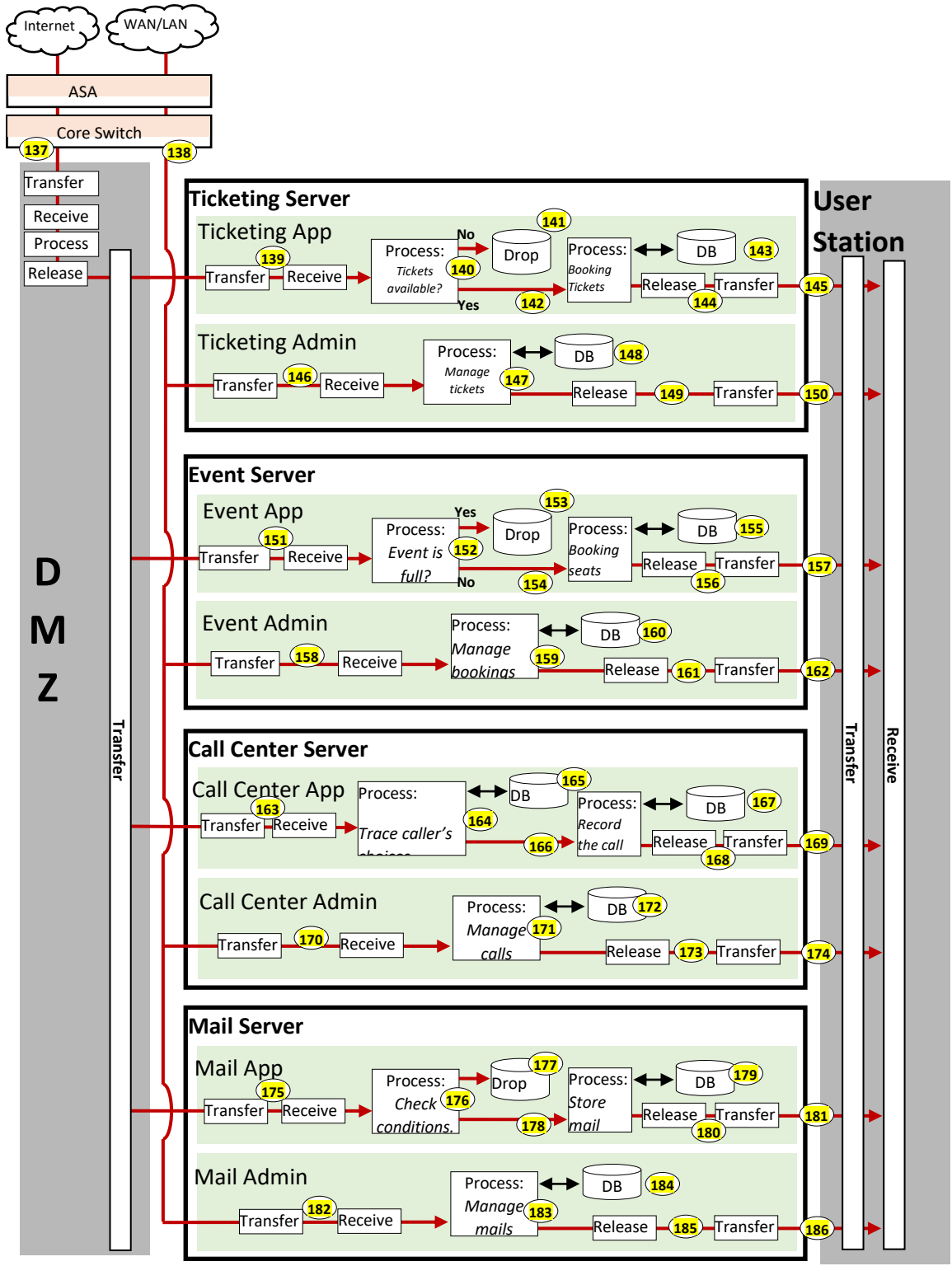

**Fig. 9:** Network architecture of internet/internal communication

*Events Server*

- For an Internet source (137), the packet flows from the DMZ to the event server. Because the packet is transferred from the DMZ, the packet arrives in the event application (151), where it is processed (152) for seat availability. If seats are unavailable, then the packet will be dropped (153). Otherwise, the packet will be processed (154) for seat booking. The packet stores (155) the information of the booking in the database. Then, the packet is released and transferred (156) back to the requester (157)

- For an internal source (138), the packet flows from the core switch to the event server. Because the packet is transferred from the core switch, the packet arrives in the event admin (158), where it is processed (159) for managing event registrations, such as for changing seat details, deleting seats and adding new categories. The new information is stored (160) in the database. Then, the packet is released and transferred (161) back to the requester (162)

*Call Center Server*

- For an Internet source (137), the packet flows from the DMZ to the call **center server.** Because the packet is transferred from the DMZ, the packet arrives in the call center application (163), where it is processed (164) to trace the caller's actions while the call center script is played and store them in the database (165). Then, the call is processed (166) for recording and stored in the database (167). Afterward, the packet is released and transferred (168) back to the requester (169) with a live phone call or a recorded message explaining that all agents are busy
- For an internal source (138), the packet flows from the core switch to the **call center server.** Because the packet is transferred from the core switch, the packet arrives at the call center admin (170), where it is processed (171) to manage calls, such as listening to calls, deleting calls and exporting call reports. The actions will either be stored in or retrieved from (172) the database. Then, the packet is released and transferred (173) back to the requester (174)

*Mail Server*

- For an Internet source (137), the packet flows from the DMZ to the mail server. Because the packet is transferred from the DMZ, the packet arrives in the mail application (175), where it is processed (176) to check mail conditions, such as ensuring that space is available for the mail and the mail is not spam or that Internet connectivity exists. If the request conflicts with any condition, the packet will be dropped (177). Otherwise, the packet will be processed (178) for storage (179) in the database. Then, the packet will be released and transferred (180) back to the requester (181) for acknowledgment that the mail has been successfully received or dropped
- For an internal source (138), the packet flows from the core switch to the mail server. Because the packet is transferred from the core switch, the packet arrives in the mail admin (182), where it is processed (183) to manage mail, such as exporting mail, adding aliases and changing licenses. The actions will either be stored in or retrieved from (184) the database. Then, the packet is released and transferred (185) back to the requester (186)

## Conclusion

In this study, we aimed to develop a new diagrammatic method, the TM model, to describe computer networks for purposes such as documenting networks to be used (e.g., as a reference in making the network's management more effective). Current description methods use either abstract graphs of nodes and edges or superficial icons connected by lines. The proposed approach is more systematic. It is based on five generic operations, which form autonomy for each node in the network with a unifying feature of flows of things among the nodes. Each item in the network is uniformly viewed as a thing that flows and a machine that handle things. Applying this method results is defining a single coherent network architecture, similar to engineering schematics. Additionally, in current depictions of network diagrams, little consideration is directed to the role of the interiority of a node, which can provide a more fine-grained understanding of networks because TM provides complete descriptions of the interiorities of the nodes participating in the network.

The advantages of TM modeling are substantiated by the description of a real organization's complex network as a case study. Further research in this direction will involve building computer-based tools to facilitate building diagrams using TM modeling.

## Authors' Contributions

All authors equally contributed in this work.

## Ethics

This article is original and contains unpublished material. The corresponding author confirms that all of the other authors have read and approved the manuscript. No ethical issues were involved and the authors have no conflict of interest to disclose.

## References

Abdullah, A.M., 2017. Application of petri nets in computer networks. Department of Applied Mathematics and Computer Science, Eastern Mediterranean University – North Cyprus.

Al-Fedaghi, S., 2019a. Five generic processes for behavior description in software engineering. Int. J. Comput. Sci. Inform. Security, 17: 120-131.

Al-Fedaghi, S., 2019b. Thing/machines (thimacs) applied to structural description in software engineering. Int. J. Comput. Sci. Inform. Security, 17: 1-11.

Al-Fedaghi, S., 2019c. Toward maximum grip process modeling in software engineering. Int. J. Comput. Sci. Inform. Security, 17: 137-156. DOI: 10.2495/SAFE-V9-N2-137-156

Al-Fedaghi, S. and M. Makdessi, 2019. Software Architecture as a Thinging Machine: A Case Study of Monthly Salary System. In: Intelligent Systems Applications in Software Engineering, Silhavy, R., P. Silhavy and Z. Prokopova (Eds.), Springer, Cham, pp: 83-97.

Al-Fedaghi, S. and O. Alsumait, 2019. Toward a conceptual foundation for physical security: Case study of an IT department. Int. J. Safety Security Eng., 9: 137-156. DOI: 10.2495/SAFE-V9-N2-137-156

Al-Fedaghi, S. and J. Al-Fadhli, 2019. Modeling an unmanned aerial vehicle as a thinging machine. Proceedings of the 5th International Conference on Control, Automation and Robotics, Apr. 19-22, Beijing, China. DOI: 10.1109/ICCAR.2019.8813706

Al-Fedaghi, S. and B. BehBehani, 2018. Thinging machine applied to information leakage. Int. J. Adv. Comput. Sci. Applic., 9: 101-110. DOI: 10.14569/IJACSA.2018.090914

Al-Fedaghi, S. and E. Haidar, 2019. Programming is diagramming is programming. J. Software, 14: 410-422. DOI: 10.17706/jsw.14.9.410-422

Al-Fedaghi, S. and M. Al-Otaibi, 2019. Service-oriented systems as a thining machine: A case study of customer relationship management. Proceedings of the IEEE International Conference on Information and Computer Technologies, Mar. 14-17, University of Hawaii, Maui College, Kahului, Hawaii, USA, pp: 243-254. DOI: 10.1109/INFOCT.2019.8710891

Bryant, L.R., 2012. Towards a machine-oriented aesthetics: On the power of art. Proceedings of the Matter of Contradiction Conference, (MCC' 12), Limousin, France.

Cisco, 2015. Cisco ASA 5500-X series firewalls, ASA 8.2: Packet flow through an ASA firewall.

Dori, D., 2002. Why significant UML change is unlikely. Commun. ACM, 45: 82-85.

Ene, M. and C. Persson, 2005. The process of process documentation-A case study at Volvo IT. M.S. Thesis, School Econ. Commercial Law, University of Gothenburg.

Farmer, J., B. Lane, K. Bourg and W. Wang, 2016. Network Documentation. In: FTTX Networks: Technology Implementation and Operation, Kaufmann, M. (Ed.), Elsevier, Amsterdam, Netherlands, ISBN-10: 0124201377.

Giagilis, G.M., G.I. Doukikidis and R.J. Paul, 1996. Simulation for intra- and inter-organizational business process modelling. Proceedings of the Winter Simulation Conference, Dec. 8-11, Baltimore, USA, pp: 1297-1304. DOI: 10.1109/WSC.1996.873439

Hassan, N.R., J. Mingers and B. Stahl, 2018. Philosophy and information systems: Where are we and where should we go? Eur. J. Inform. Syst., 27: 263-277.

Heidegger, H., 1975. The Thing. In: Poetry, Language, Thought, Hofstadter, A. (Ed.), Harper and Row, New York, pp: 161-184.

Kinghorn, G., 2018. A better way to document your network topology than Visio. Forward Networks.

Olzak, T., 2006. A practical approach to threat modeling. Erudio Security, LLC., Toledo, OH.

Portal Systems, 2019. Network documentation.

Riemer, K., R.B. Johnston, D. Hovorka and M. Indulska, 2013. Challenging the philosophical foundations of modeling organizational reality: The case of process modeling. Proceedings of the International Conference on Information Systems, (CIS' 13), Milan, Italy.

Yejian Technologies, 2020. Types of networks: Main 5 types of computer networks, Router-switch.com Site.

Ryom, C., 2019. Implementing the Cisco Adaptive Security Appliance (ASA) [PPT Presentation], SlideServe.

Sherman, J., 2019. Network documentation 101. Graphican Networks.

SmartDraw, 2019. Network diagram.

Sopwith, T., 2019. How to document: Network. ITGlue.

Wolf, T., J. Griffioen, K. Calvert, R. Dutta and G. Rouskas et al., 2012. Choice as a principle in network architecture. Proceedings of the ACM Conference on Applications, Technologies, Architectures and Protocols for Computer Communication, Aug. 13-17, ACM, Helsinki, Finland.